\documentstyle[prd,aps,twocolumn,amstex,epsfig]{revtex}

\newcommand{\ket}[1]{|#1\rangle}

\newcommand{\bea}{\begin{eqnarray}}
\newcommand{\ea}{\end{eqnarray}}
\newcommand{\ord}{{\mathcal{O}}}

\begin{document} 

\wideabs{ 
\title{Pattern recognition on a quantum computer}
\author{Ralf Sch\"utzhold} 
\address{
Department of Physics and Astronomy,
University of British Columbia,
Vancouver B.C., V6T 1Z1 Canada
\\
Electronic address: {\tt schuetz@@physics.ubc.ca}
}
\date{\today}

\maketitle

\begin{abstract} 
By means of a simple example it is demonstrated that the task of finding 
and identifying certain patterns in an otherwise (macroscopically) 
unstructured picture (data set) can be accomplished efficiently by a 
quantum computer.
Employing the powerful tool of the quantum Fourier transform the proposed
quantum algorithm exhibits an exponential speed-up in comparison with its 
classical counterpart.
The digital representation also results in a significantly higher accuracy 
than the method of optical filtering.
\\
PACS: 
03.67.Lx, 
03.67.-a, 
42.30.Sy, 
89.70.+c. 
\end{abstract} 
}

\section{Introduction}\label{Introduction}
\begin{figure}[ht]
\centerline{\mbox{\epsfxsize=8cm\epsffile{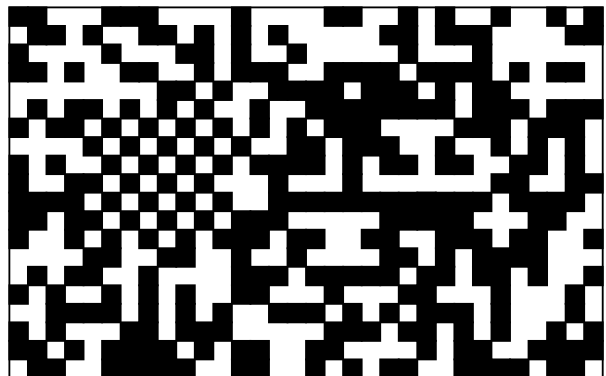}}}
\caption
{A $32\times20$ array half-filled with points, i.e.~$\varrho=1/2$.
In an $8\times8$ square they form a pattern (hence $\chi=1/10$), 
otherwise the points are randomly distributed.}
\label{pattern}
\end{figure}

%
Pattern recognition is one of the basic problems in artificial intelligence, 
see e.g.~\cite{fukunaga}. 
For example, generally a short look at a picture like the one in 
Fig.~\ref{pattern} suffices for the human brain to spot the region with 
the pattern. 
However, it is a rather non-trivial task to accomplish the same performance
with a computer -- in particular if the orientation and the structure of the 
pattern are not known {\em a priori} and it is not perfect (in contrast to 
the one in Fig.~\ref{pattern}).

Besides the detection and localisation of pattern 
(for example identifying seismic waves in the outputs of seismographs)
the comparison and matching of the observed pattern to a set of templates 
(such as face recognition) is another interesting question.
Usually these problems are solved with special classifiers, such as neuronal
networks or Fourier analysis, etc., cf.~\cite{fukunaga}. 

The specific properties of the task of pattern recognition 
(one may consider many combinations simultaneously and is interested in 
global features only) give raise to the hope that quantum algorithms 
may be advantageous in comparison with classical (local) computational 
methods (with a unique entry).

During the last decade the topic of quantum information processing has 
attracted increasing interest, see e.g.~\cite{nielsen} for a review.
It has been shown that quantum algorithms can be enormously faster than the 
best (known) classical techniques:
Shor's factoring algorithm \cite{shor}, which exhibits an 
exponential speed-up relative to the best known classical method;
Grover's search routine \cite{grover} with a quadratic speed-up;
and several black-box problems 
\cite{deutsch,deutsch-jozsa,simon,bernstein-vazirani},
some of which also exhibit an exponential speed-up; etc.

In the following a quantum algorithm for the detection, identification, 
and localisation of certain patterns in an otherwise (macroscopically) 
unstructured data set is presented.
It turns out that this method is exponentially faster than its classical
counterpart too.
Furthermore, it outperforms the (also extremely fast) method of optical 
filtering in terms of accuracy.

The advantages of using quantum memories and computers for the 
aforementioned task of template matching 
(which is somewhat different from pattern detection/localisation) 
have been discussed in \cite{memory,template}.
Note, however, that the necessity of loading the complete data set into a 
quantum memory may represent a drawback (cf.~the next Sections).
In Ref.~\cite{cluster} an algorithm for data clustering 
(in pattern recognition problems) is developed, which is based on/inspired by
principles of quantum mechanics -- but does not involve quantum computation.

%
\section{Description of the Problem}\label{Description}
%
Let us consider a rectangular $N \times M$ array of unit cells which are 
either absorptive (black) or reflective/transparent (white), 
cf.~Fig.~\ref{pattern}.
In the following we shall assume that these cells are perfectly absorbing
or reflecting/transmitting, respectively, but reasonable deviations from this
idealised behaviour resulting in finite absorption, reflection, and 
transmission coefficients do not alter the following considerations.

The white cells are distributed with a roughly homogeneous density $\varrho<1$
(for example $\varrho=1/2$) and will later be called points for brevity, 
i.e., the $N \times M$ array contains $P=\varrho NM$ points.
A small fraction $\chi$ of these points (say $\chi=1/10$) forms a pattern
in a connected (but not necessarily rectangular) region of the size $\chi NM$, 
cf.~Fig.~\ref{pattern}.

In contrast to Fig.~\ref{pattern} the pattern does not need to be perfect -- 
average symmetries are sufficient, see Section~\ref{Pattern} below.
For simplicity we restrict our consideration to linear 
(but again not necessarily rectangular) patterns.
I.e., the patterns are supposed to be (approximately) invariant under at least
two symmetry transformations described by global translations into different 
directions.
Geometrically speaking, the angles within the pattern do not change, 
think of, for example, a set of parallel and equidistant lines or a 
periodic repetition of small elements as in some wall papers, etc.

Let us suppose that we know $N$, $M$, and $\varrho$, for these quantities
can easily be measured having at hand the $N \times M$ array of unit cells 
(which might be some sensitive nano-device, for example) -- 
but we neither know the size $\chi$ of the pattern 
(or whether there is a pattern at all) nor its structure and orientation
(except that it is linear).
The task is to find an algorithm for extracting this information.

In principle, we may find out the position of all points (white cells)
by shining appropriately focused light beams on the array and measuring 
the reflection or transmission.
However, the array is also assumed to be very sensitive and each absorbed
photon causes a certain amount of damage -- similar to the Elitzur-Vaidman 
problem (detection without destruction), cf.~\cite{bomb}.
Therefore, the number of incident photons should be as small as possible.

At the same time, we wish to obtain the characteristic parameters of the 
pattern with maximally attainable accuracy and minimal effort 
(i.e.~number of subsequent operations). 
In order to cast the above requirements into a well-defined complexity 
theoretic form we consider the limit of very large arrays $N,M\gg1$.
Furthermore, we assume $N=2^n$ and $M=2^m$ with integers $n,m\in{\mathbb N}$ 
allowing for a binary representation.
(Otherwise we may enlarge the array accordingly or consider a part of it only.)
%
\section{Data Processing}\label{Data}
%
It is probably most convenient to view the data set as a (quantum) black box
\bea
\label{bb}
{\mathcal{BB}}
\; : \;
\left[
\begin{array}{c}
\ket{x}\\
\ket{y}\\
\ket{0}
\end{array}
\right]
\; \rightarrow \;
\left[
\begin{array}{c}
\ket{x}\\
\ket{y}\\
\ket{f(x,y)}
\end{array}
\right]
\,,
\ea
where the input state encodes the coordinates $x$ and $y$ of a potential point
(a white cell) in the array as $n$- and $m$-qubit strings, respectively, 
together with a third one-qubit register $\ket{0}$ needed for unitarity.
The output function  $f(x,y)$ assumes the value 1 if there is a point at these
coordinates and 0 if not.

As a possible physical realization one might imagine a configuration like the 
following:
a focused light beam passes $n+m$ controlled refractors 
(e.g.~non-linear Kerr media) which change
its direction by definite angles $\varphi_j$ if the control qubit is $\ket{1}$ 
and do not affect it otherwise.
For suitably chosen angles 
($\varphi_j^{x,y}=\varphi_0^{x,y}/2^j$ with $j\in{\mathbb N}$)
the final direction of the beam encodes the position $(x,y)$ on the 
$N \times M$ array if the digits of the coordinates 
$\ket{x},\ket{y}$ are inserted as the control qubits.   
Shining the so directed light beam (which may consist of only one photon)
through an aperture mask like Fig.~\ref{pattern} on a detector reproduces 
the action of the black box in Eq.~(\ref{bb}).

Note that, in this realization, it is not necessary to load the complete 
(classical) information of the array into a quantum memory.
This would slow down the whole process drastically and hence represent 
a serious caveat. 
(The same problem limits the region of applicability of Grover's quantum 
search procedure, for example.)
In addition, it is hard to see how this could be done without shining a 
relatively large number of photons on the array and thereby destroying it,
cf.~Section~\ref{Description}.

Moreover, the involved number ($n+m$) of devices (refractors) is very small 
in this case.
Each refractor acts roughly similar to a controlled swap or a switch gate
\bea
\label{sw}
{\mathcal{SW}}
\; : \;
\left[
\begin{array}{c}
\ket{\alpha}\\
\ket{\beta}\\
\ket{0}
\end{array}
\right]
\; \rightarrow \;
\left[
\begin{array}{c}
\ket{\alpha}\\
\ket{\neg\alpha\wedge\beta}\\
\ket{\alpha\wedge\beta}
\end{array}
\right]
\,,
\ea
with $\alpha,\beta=0,1$; but the series connection of these refractors 
allows for a very efficient data processing.
%
\section{Quantum Algorithm}\label{Quantum}
%
Now we may apply the well-known trick of inquiring all possible values of the 
coordinates $(x,y)$ in only one run of the black box (quantum parallelism). 
To this end we prepare a state as the superposition of all possibilities
by using the Hadamard gate ${\mathcal{H}}$.
For a single qubit in the $\ket{0},\ket{1}$ basis the (unitary) Hadamard gate 
${\mathcal{H}}$ acts as ${\mathcal{H}}\ket{0}=(\ket{0}+\ket{1})/\sqrt{2}$. 
By multiple application of ${\mathcal{H}}$ and running the black box 
(only once) we arrive at the desired superposition
\bea
\label{super}
{\mathcal{BB}}
\left[
\begin{array}{c}
{\mathcal{H}}^{(n)}\ket{0^{(n)}}\\
{\mathcal{H}}^{(m)}\ket{0^{(m)}}\\
\ket{0}
\end{array}
\right]
=
\frac{1}{\sqrt{NM}}\sum\limits_{x=0}^{N-1}\sum\limits_{y=0}^{M-1}
\left[
\begin{array}{c}
\ket{x}\\
\ket{y}\\
\ket{f(x,y)}
\end{array}
\right]
\,.
\ea
Now measuring the third register $\ket{f}$ and obtaining 1 prepares 
the state $\ket{\Psi}$ as a superposition of the coordinates $\ket{x}$
and $\ket{y}$ of all points.
Assuming an ideal black box, the outcome $f=0$ would just imply the 
complementary set $\varrho\rightarrow1-\varrho$.
However, in the presence of absorbing units as described in
Section~\ref{Description}, the resulting entanglement with the state of the 
absorptive cells would destroy the coherence completely in the case $f\neq1$. 
  
It will be advantageous to reorganise the array by dividing it into
$M$ rows of length $N$ and combining them all to one string of length $S=NM$.
The coordinate of a given point is now one $n+m=s$-digit binary number $z=x+Ny$
(instead of two numbers $x$ and $y$).
The corresponding quantum state is simply given by 
$\ket{z}=\ket{x}\otimes\ket{y}$.
In terms of this representation the quantum state $\ket{\Psi}$ prepared by 
the measurement of $f=1$ reads
\bea
\label{measure}
\ket{\Psi}=\frac{1}{\sqrt{\varrho S}}\sum\limits_{l=1}^{\varrho S}\ket{z_l}
\,,
\ea
where $0 \geq z_l \geq S-1$ denotes position of $l$-th point 
(as a $s$-digit binary number).

The next basic part of the quantum algorithm is the application of the 
quantum Fourier transform (QFT). 
It acts on a basis element like $\ket{z}=\ket{110100\dots}$ as
\bea
\label{def}
{\mathcal{QFT}}
\; : \;
\ket{z}
\; \rightarrow \;
\frac{1}{\sqrt{S}}\sum\limits_{k=0}^{S-1}
\exp\left(2\pi i\,\frac{zk}{S}\right)\ket{k}
\,.
\ea
Hence the superposition state $\ket{\Psi}$ in Eq.~(\ref{measure}) 
will be transformed into
\bea
\label{qft}
{\mathcal{QFT}}\ket{\Psi}
=\sum\limits_{k=0}^{S-1}\sum\limits_{l=1}^{\varrho S}
\frac{1}{S\sqrt{\varrho}}\exp\left(2\pi i\,\frac{z_lk}{S}\right)\ket{k}
\,.
\ea
Assuming a distribution of points $z_l$ without any (macroscopic) pattern 
(e.g.~purely random) there will be no privileged values of $k$ 
(except $k=0$) and the measurement of $k>0$ yields just noise.
However, the presence of a pattern within the data set introduces
a typical length scale and thus leads to peaks of the factor in front of
$\ket{k}$ at certain values of $k$ -- which hence can be used as an indicator,
see the next Section.
In this way the QFT efficiently solves the problem of feature selection -- 
i.e., extracting a small amount of relevant quantities (such as wave-numbers) 
from a large data set.
%
\section{Pattern localisation}\label{Pattern}
%
The task of pattern recognition does not only include the mere detection of
a pattern but also its localisation and classification.
The comparison with a given set of templates will not be discussed here,
see e.g.~\cite{memory,template}.
The next step is to extract informations about the pattern from the peaks
in the measurements of $k$ -- in close analogy to the reconstruction of the
probe structure from the Laue diagram in diffraction experiments.

Consider, for example, a simple pattern consisting of parallel lines like 
the one in Fig.~\ref{pattern}.
In this case the basic quantities are the distance $D=\rm const$ of the lines 
as well as their orientation as described by the (constant) angle 
$-\pi/2\leq\vartheta\leq\pi/2$.
Here ${\vartheta}$ denotes the deviation of the line from a vertical one,
i.e., after going down $R$ rows the sequence is shifted by 
${\tan\vartheta}\,R$ columns to the right.
So the points $z$ marking the centre of a particular line are given by
\bea
\label{line}
z=z_0+[{\mathbb N}(N+{\tan\vartheta})]_{\rm integer}
\,.
\ea
Note that (in contrast to Fig.~\ref{pattern}) the lines do not need to be
perfect -- it is sufficient if, in average, the density of points within a 
line-width of, say, $D/2$ deviates by a finite amount $\Delta\varrho$ 
(e.g.~$\Delta\varrho=1/4$) from the mean $\varrho$.

According to Eq.~(\ref{qft}) every row of the pattern generates peaks at
\bea
\label{row}
k=\left[
{\mathbb N}\,\cos\vartheta\,\frac{S}{D}\pm
\ord\left(\frac{M}{D\sqrt{\chi}}\right)
\right]_{\rm integer}
\,,
\ea
with the second term denoting their width.
Both, the position and the width of the peaks can be obtained from
the associated Laue function 
${\mathfrak L}(\xi,\kappa)=\sin^2(\pi\xi\kappa)/\sin^2(\pi\kappa)$
with $k=\kappa\cos\vartheta\,S/D$ and $\xi=\ord(N\sqrt{\chi})$
in this case.

However, the sum of all rows interferes constructively only if $k$ is 
fine-tuned according to
\bea
\label{column}
k=\left[
{\mathbb N}\,\frac{N-\tan\vartheta}{N}\,M\pm
\ord\left(\frac{1}{\sqrt{\chi}}\right)
\right]_{\rm integer}
\,,
\ea
which again can be obtained from the associated Laue function with now
$\kappa=k(N+\tan\vartheta)/S$ and $\xi=\ord(M\sqrt{\chi})$.
The strongest peaks in the measurements of $k$ occur for values which 
satisfy both conditions (\ref{row}) and (\ref{column}) simultaneously.
Accordingly, the wave-numbers of those potential peaks read
\bea
\label{resonance}
k\approx\left[{\mathbb N}\left(
\cos\vartheta\,\frac{S}{D}-\sin\vartheta\,\frac{M}{D}
\right)\right]_{\rm integer}
\,,
\ea
where the corresponding width/uncertainty has been omitted.

However, for large enough $D$, not every peak in Eq.~(\ref{row}) will contain 
wave-numbers matching Eq.~(\ref{column}) in general.
The condition for this to happen is that the integer ($\mathbb N$) in 
Eq.~(\ref{row}) multiplied by $\cos\vartheta\,N/D$ is again close to 
another integer -- the one in Eq.~(\ref{column}) -- within an accuracy of
$\ord(1/[D\sqrt{\chi}])$.

Therefore, not all the $k$-values in Eq.~(\ref{resonance}) do necessarily
represent large peaks -- the first few of them may be suppressed.
On the other hand, the larger $D$ is the more $k$-values of potential peaks 
in Eq.~(\ref{resonance}) are contained in the interval $0<k<S$.
Consequently, from the number $D/\cos\vartheta$ of potential peaks in
Eq.~(\ref{resonance}) there must be at least a few $\ord(1/\sqrt{\chi})$
which match both conditions (\ref{row}) and (\ref{column}).

Determining the largest common factor of all the wave-numbers of the peaks
(within the given accuracy) we obtain a value for the expression in  
Eq.~(\ref{resonance}) or integer multiples of it.
Unfortunately, this information alone is not sufficient for extracting 
$D$ and $\vartheta$.
To this end we may simply transpose the array (e.g.~Fig.~\ref{pattern}) 
by interchanging rows and columns $N \leftrightarrow M$ and run the same
algorithm again.
Since transposing corresponds to $\vartheta\to\pi/2-\vartheta$ the 
wave-numbers of the peaks are now 
\bea
\label{trans-resonance}
k'\approx\left[{\mathbb N}\left(
\sin\vartheta\,\frac{S}{D}-\cos\vartheta\,\frac{N}{D}
\right)\right]_{\rm integer}
\,.
\ea
Combining the possible values for $D/\cos\vartheta$ from Eq.~(\ref{resonance})
with the ones for $D/\sin\vartheta$ from Eq.~(\ref{trans-resonance}) we obtain
approximate candidates for $D$ and $\vartheta$.
Comparison with the fine-tuning conditions such as Eqs.~(\ref{row}) and 
(\ref{column}) as well as the knowledge of which peaks are suppressed and 
which are not allows us to extract the actual values of $D$ and $\vartheta$
with high (in fact, maximally attainable) precision $\ord(1/\sqrt{\chi})$.

The height of the peaks can be estimated by means of Eq.~(\ref{qft}).
In the resonance case the sum includes $S\,\chi\,\Delta\varrho$ 
constructively interfering addends which lead to an amplitude of order
$\ord(\chi\,\Delta\varrho/\sqrt{\varrho})$.
Thus the probability $p$ of measuring the peaks in 
Eqs.~(\ref{resonance}) and (\ref{trans-resonance}) is given by
\bea
\label{prob}
p=\ord\left(\frac{(\chi\,\Delta\varrho)^2}{\varrho}\right)\,,
\ea
i.e.~independent of $N$ and $M$ -- and therefore drastically enhanced 
over the (random) noise.

Consequently, if a number $\Omega$ of measurements yields one or more 
pronounced peaks besides $k=0$ then there exists a pattern larger than
$\chi_{\rm min}=\ord(1/\sqrt{\Omega})$ and otherwise not
(at least with a very high probability).
Quite reasonably, the smaller the pattern, i.e.~$\chi$, 
the longer one has to search.

After having solved the feature selection problem efficiently by the quantum
algorithm the remaining analysis (peak finding and stopping criteria, etc.) 
of a small (independent of $N$ and $M$) amount of measured wave-numbers 
can be accomplished by a classical algorithm.

In this way one can determine the size of the pattern $\chi$ by the
frequency of measuring the peaks at $k$ and $k'$ (and their width).
Its structure, i.e.~the values of $D$ and $\vartheta$, 
can be inferred from the location of the peaks. 

Having found the parameters $D$, $\vartheta$, and $\chi$ of the pattern 
it can be localised easily -- for example by dividing the total $N \times M$ 
array into smaller pieces (according to $\chi$) 
and running the same quantum algorithm again in the smaller domains. 

More complicated (but still linear) patterns, such as a regularly recurring 
pictures\footnote{Again, these pictures do not need to be perfect -- 
average features are sufficient.}
(like in many wall papers), possess more than one characteristic angles
$\vartheta$ in general and therefore generate a richer peak structure -- 
but the main idea remains the same.
%
\section{Complexity analysis}\label{Complexity}
%
Let us estimate the size of the proposed algorithm, i.e.~the number of 
involved computational steps, and compare it with the classical method,
in the limit $S\to\infty$ while $\varrho$, $\Delta\varrho$, and $\chi$ 
remain finite.

In view of Eq.~(\ref{prob}) we need only a few $\ord(S^0)$ queries of the
black box in order to find a pattern of a given size with high probability.
Clearly, this is not possible with any classical algorithm -- demonstrating 
the advantage of the global quantum computation over the local 
(only one point at the time) classical technique.
Since the number of queries of the black box corresponds to the total amount 
of photons shining on the array, the quantum algorithm causes less damage
(cf.~Section~\ref{Description}) than any classical method, 
see also~\cite{bomb}.

Given the explicit physical realization of the black box described in 
Section~\ref{Data} 
it is also possible to compare the total number of fundamental manipulations.
For the preparation of the initial state in Eq.~(\ref{super}) one has to
apply the Hadamard gate $m+n=s=\log_2S$ times.
The black box itself involves about the same number of operations. 
The quantum Fourier transform (QFT) in Eq.~(\ref{def}) requires 
$\ord(\log_2^2S)$ steps for obtaining the exact result and is even faster 
$\ord(\log_2S)$ if we measure \cite{nielsen} the outcome immediately 
afterwards -- as it is the case here.

In contrast, the best known classical algorithm, the fast Fourier transform
(FFT), implements $\ord(S\log_2S)$ operations and is therefore exponentially
slower.
Note that, since we do not know the typical ``wave-numbers'' $k$ associated 
with the pattern {\em a priori}, we would have to calculate the FFT for a 
large number $\ord(S)$ of possible values of $k$ -- whereas the QFT 
accomplishes all this simultaneously and automatically gives us the values 
$k$ with the largest amplitudes in average measurements.

Unfortunately, it cannot be excluded here that perhaps a classical algorithm 
exists which is better than the FFT and may compete with the proposed quantum 
algorithm (though not in the number of queries of the black box).
But since the processing of the coordinates of only one single point already 
requires $\ord(\log_2 S)$ operations, one would have to find the pattern
by considering a few $\ord(S^0)$ points in order to outrun the quantum computer
-- which is apparently not possible.

Nevertheless, in certain situations -- e.g.~for perfect lines with $D=\ord(1)$
such as in Fig.~\ref{pattern} -- it is possible to design an appropriate 
classical algorithm which determines $\vartheta$ by using only 
$\ord(\log_2^q S)$ points with $q\geq1$, which requires 
$\ord(\log_2^{q+1} S)$ computational steps.
In this case the speed-up is merely polynomial.

In the more general case, however, where $D$ can be very large $D\gg1$ and 
the lines are not perfect, it is really hard to see how one might 
extract basically the same information as the FFT in (polynomially) 
logarithmic time classically. 
(An $\ord(\sqrt{S})$ algorithm, for example, would also be exponentially 
slower than the QFT.)

Assuming that there is indeed no such classical algorithm
(or, that the set of patterns for which the Fourier transform is the best
classifier is not empty) the problem under consideration represents another 
example for the (conjectured) exponential speed-up of quantum computation -- 
based on the power of QFT for problems related to (quasi) periodical 
structures (which is also the basic ingredient for Shor's algorithm 
\cite{shor}; though in that case the periodicity is exact -- 
in contrast to the situation considered here).

Of course, this speed-up has only been possible since it was not necessary
to load the complete array into a quantum memory 
(cf.~\cite{memory,template}) -- this would have involved about $\ord(S)$ 
operations and thereby lead to a drastic (exponential) slow-down.
%
\section{Summary and Outlook}\label{Summary}
%
In summary, quantum algorithms are capable of solving certain problems of 
pattern recognition (i.e.~detection, localisation, and classification) 
besides template matching \cite{memory,template} much faster 
$\ord(\log_2 S)$ than their classical counterparts \cite{speculation}.

Although this has been demonstrated explicitly in this article for line 
patterns only, the basic idea applies to more difficult (but still linear) 
patterns as well.
(In some sense, this idea arises from the often successful approach to copy
the elegant solutions which nature reveals to us -- such as the possibility of
distinguishing a crystal from an amorphous material via X-ray diffraction
or the method of optical filtering, see the next Section.)  
The investigation of non-linear patterns, such as a set of concentric circles,
is apparently more involved and probably requires adapted methods.

Together with the findings in Refs.~\cite{memory,template} the results of 
the present article give raise to the hope that quantum algorithms are 
also advantageous for more general pattern recognition problems.
%
\section{Optical filtering}\label{Optical}
%
Interestingly, the manipulations involved in the proposed quantum algorithm
-- apart from the physical realization of the black box itself -- can be 
accomplished (at least in principle) with present-day optical devices:
Hadamard gates as well as the calculation of the quantum Fourier transform
with subsequently measuring the outcome can be realized using beam splitters
and classically controlled phase shifters, cf.~\cite{nielsen}.
This observation leads to the question of whether one could achieve 
a similar performance with purely optical techniques.

Indeed, the method of optical filtering reproduces some key features of the 
proposed quantum algorithm.
Shining a plane wave with an appropriate wave-number on an aperture mask
like Fig.~\ref{pattern} the far field diffraction amplitudes are given by
the Fourier transform of the object (e.g.~Fig.~\ref{pattern}) in terms of
the perpendicular wave-number.
Using an ordinary convex lens one may convert these wave-numbers into
positions in the focal plane of the lens, see e.g.~\cite{optics}.

In this way the described apparatus effectively calculates the desired 
Fourier transform.
However, achieving the same accuracy as the proposed quantum algorithm,
i.e.~$\ord(\log_2 S)$ digits for $\vartheta$ and $D$, would require 
exponential precision and hence exponential effort: 
number of detectors in the focal plane, etc.
Another problem is the fact that $D$ is not known {\em a priori} and may 
vary over several orders of magnitude -- which makes it difficult to select
a suitable wave-number for the incident light.

These obstacles are caused by the main difference between optical filtering 
and the proposed quantum algorithm.
In optical filtering, the relevant quantities 
(such as wave-number, position in the focal plane, $\vartheta$, and $D$) 
are directly, i.e.~linearly, related to each other -- 
whereas the quantum algorithm employs the digital representation
and therewith outperforms the former method.

On the other hand, if we happen to know the order of magnitude of $D$ in
advance and would be willing to settle for a limited accuracy of only a 
few $\ord(S^0)$ digits, we may detect the pattern with optical filtering
by using only a few $\ord(S^0)$ photons.
Although much less powerful, i.e.~accurate, this method would be significantly
faster than the proposed quantum algorithm, which also uses only a few 
photons shining on the pattern -- but it requires $\ord(\log_2 S)$ photons
(or, more generally, qubits) for the subsequent data analysis (QFT, etc.).
(This enhancement in speed -- although at a certain cost -- reflects the fact 
that a quantum field theoretical object such as a photon has more degrees of 
freedom than just one single qubit.)
%
\section{Appendix}
%
If there were no absorptive cells in the array at all but only (perfectly) 
reflective (white) and transparent (black) ones, one could (in principle) 
realize the following generalisation of the black box in Eq.~(\ref{bb})
\bea
\label{bb-g}
{\mathcal{BB}}
\; : \;
\left[
\begin{array}{c}
\ket{x}\\
\ket{y}\\
\ket{\alpha}
\end{array}
\right]
\; \rightarrow \;
\left[
\begin{array}{c}
\ket{x}\\
\ket{y}\\
\ket{\alpha\oplus f(x,y)}
\end{array}
\right]
\,,
\ea
with $\alpha=0,1$. Here $\oplus$ denotes summation modulo 2,
i.e., $1\oplus0=0\oplus1=1$ and $0\oplus0=1\oplus1=0$.
In this case one may improve the quantum algorithm by sending the
superposition state in Eq.~(\ref{super}) with the third register 
being $(\ket{0}-\ket{1})/\sqrt{2}$ instead of $\ket{0}$ to the black box.
The third register does not change during this procedure and the resulting
state encodes the information about the points in the array in the 
phases ($+1$ or $-1$) instead of the amplitudes ($1$ or $0$)
\bea
\label{sign}
\ket{\Psi}
=
\frac{1}{\sqrt{S}}\sum\limits_{z=0}^{S-1}
(-1)^{f(z)}\,\ket{z}
\,.
\ea
Assuming $\varrho=1/2$, the quantum Fourier transform of this state
has certain advantages over the one in Eq.~(\ref{measure}) -- one gets rid of
the (useless) peak at $k=0$ and enhances the probabilities of the other peaks 
by a factor of 2.
Unfortunately, it is hard to see how one might be able to exploit this
advantage of the modified black box in the presence of absorbing units.
%
\section*{Acknowledgements}
%
The author acknowledges valuable discussions with L.~D'Afonseca, D.~Curtis,
S.~Fuchs, Y.~Gusev, D.~Meyer, N.~Pippenger, S.~Popescu, G.~Schaller, 
C.~Trugenberger, and B.~Unruh.
This work was supported by the Alexander von Humboldt foundation and by the 
Natural Science and Engineering Research Council of Canada.
%
\addcontentsline{toc}{section}{References}


\begin{thebibliography}{99}
\bibitem{speculation}
Speculating a little outside the scope of this article one might ask whether
the human brain, which is capable of solving such pattern recognition problems
very well, actually makes use of similar quantum effects.
Although this idea has some appealing aspects 
(think of optical illusions, for example) it appears to be somewhat 
unlikely -- at least at a first glance.
\end{thebibliography}
\end{document}